\documentclass[a4paper,twoside,10pt]{letter}
\usepackage{graphicx,saj,multicol,subeqnarray}

\def\papertype{Preliminary report}

\def\udc{...}
\begin{document}

\baselineskip=3.1truemm
\columnsep=.5truecm
\newenvironment{lefteqnarray}{\arraycolsep=0pt\begin{eqnarray}}
{\end{eqnarray}\protect\aftergroup\ignorespaces}
\newenvironment{lefteqnarray*}{\arraycolsep=0pt\begin{eqnarray*}}
{\end{eqnarray*}\protect\aftergroup\ignorespaces}
\newenvironment{leftsubeqnarray}{\arraycolsep=0pt\begin{subeqnarray}}
{\end{subeqnarray}\protect\aftergroup\ignorespaces}
%

% Running titles

\markboth{\eightrm ON THE TIMESCALE FORCING IN ASTROBIOLOGY}
{\eightrm B. VUKOTI{\' C} and M.M. {\' C}IRKOVI{\' C}}

{\ }

\publ

\type

{\ }

% Title

\title{ON THE TIMESCALE FORCING IN ASTROBIOLOGY}

% Authors

\authors{B. Vukoti{\' c}\footnote{Corresponding author: {\tt
bvukotic@aob.bg.ac.yu}.} and M. M. {\' C}irkovi{\' c}}

\vskip3mm

% Address

\address{Astronomical Observatory Belgrade,\break Volgina 7, 11160 Belgrade-74, Serbia}

% Received and Accepted dates

\dates{May 15, 2007}{In press}

\vspace{0.7cm}

\summary{We investigate the effects of correlated global regulation
mechanisms, especially Galactic gamma-ray bursts (GRBs), on the
temporal distribution of hypothetical inhabited planets, using
simple Monte Carlo numerical experiments. Starting with recently
obtained models of planetary ages in the Galactic Habitable Zone
(GHZ), we obtain that the times required for biological evolution
on habitable planets of the Milky Way are highly correlated.
These results run contrary to the famous anti-SETI anthropic
argument of Carter, and give tentative support to the ongoing and
future SETI observation projects.}


\begin{multicols}{2}
{

\section{1. INTRODUCTION}

Recent astrobiological developments have helped elaborate the
concept of the Galactic Habitable Zone (henceforth GHZ), as the
region of the Galaxy containing habitable planets. The exact
boundaries of GHZ are still uncertain, although the basic
physical processes determining it are clear: build-up of
metallicity through the Galactic chemical evolution, frequency of
close stellar encounters and supernovae, and, possibly, the
cosmogonical effects of environmental UV irradiation. In general,
GHZ has the form of an annular ring, several kpc wide, and
comprising the Solar circle at galactocentric distance of 8.5
kpc. A great leap forward occurred with the work of Lineweaver
and his collaborators (Lineweaver 2001; Lineweaver, Fenner and
Gibson 2004) on the age distribution of planets in GHZ. One of
the most interesting consequences of that study is that the
median age of terrestrial planets in the Milky Way is $1.8 \pm
0.9$ Gyr greater than the age of the Earth (a finding making the
classical Fermi paradox even more disturbing!\footnote{For fine reviews see
Brin (1983); Duric and Field (2003).}). In the meantime,
advances in evolutionary biology and paleontology have recently
reaffirmed the determining role of mass extinction episodes in the
determining the outcome of evolution of the biosphere on Earth.
This offers a useful framework for trying to access how likely
the completion of biological evolution (in the sense of Carter's
1983 paper; see below) is in the wider context of GHZ. It is our
contention that the best approach to this problem lies with large-scale
numerical simulations which could be updated with any improvement in our
understanding of the underlying astrophysical and astrochemical mechanisms.
In a way, this is analogous to using Monte Carlo simulations in other
branches of physics where the detailed knowledge of individual subsystems'
history and properties is unobtainable or undesirable, and only the
global outcome subject to specific boundary conditions is of interest
(e.g. percolation or diffusion models).

\section{2. SIMPLE MODEL}

An important paper of Annis (1999) opened a new vista by
introducing (though not quite explicitly) the notion of global
regulation mechanism, that is, a dynamical process preventing or
impeding uniform emergence and development of life all over the
Galaxy.  In Annis' model, which he dubbed the phase-transition
model for reasons to be explained shortly, the role of such
global Galactic regulation is played by gamma-ray bursts
(henceforth GRBs ), colossal explosions caused either by terminal
collapse of supermassive objects ("hypernovae") or mergers of
binary neutron stars. GRBs observed since 1950s have been known
for more than a decade to be of cosmological origin.
Astrobiological and ecological consequences of GRBs and related
phenomena have been investigated recently in several studies
(Thorsett 1995; Dar 1997; Scalo and Wheeler 2002; Thomas et al.
2005). To give just a flavor of the results, let us mention that
Dar (1997) has calculated that the terminal collapse of the
famous supermassive object Eta Carinae could deposit in the upper
atmosphere of Earth the energy equivalent to the simultaneous
explosions of 1 kiloton nuclear bomb per km$^2$ all over the
hemisphere facing the hypernova! According to the calculations of
Scalo and Wheeler (2002), a Galactic GRB can be lethal for
eukaryotes up to the huge distance of 14 kpc. Thus, this "zone of
lethality"  for advanced lifeforms is bound to comprise the
entire GHZ whenever a GRB occurs within inner 10 kpc of the
Galaxy. Annis suggested that GRBs could cause mass extinctions of
life all over the Galaxy (or GHZ), preventing or arresting the
emergence of complex life forms. Thus, there is only a very small
probability that a particular planetary biosphere could evolve
intelligent beings in our past. However, since the regulation
mechanism exhibits secular evolution, with the rate of
catastrophic events decreasing with time, at some point the
astrobiological evolution of the Galaxy will experience a change
of regime. When the rate of catastrophic events is high, there is
a sort of quasi-equilibrium state between the natural tendency of
life to spread, diversify, and complexify, and the rate of
destruction and extinctions. When the rate becomes lower than
some threshold value, intelligent and space-faring species can
arise in the interval between the two extinctions and make
themselves immune (presumably through technological means) to
further extinctions.

It is important to understand that the GRB-mechanism is just one
of possible physical processes for "resetting astrobiological
clocks". Any catastrophical mechanism operating (1) on
sufficiently large scales, and (2) exhibiting secular evolution
can play a similar role. There is no dearth of such mechanisms;
some of the bolder ideas proposed in literature are cometary
impact-causing "Galactic tides" (Asher et al. 1994; Rampino
1997), neutrino irradiation (Collar 1996), clumpy cold dark
matter (Abbas and Abbas 1998), or climate changes induced by
spiral-arm crossings (Leitch and Vasisht 1998; Shaviv 2002).
Moreover, all these effects are cumulative: the total risk
function of the global regulation is the sum of all risk
functions of individual catastrophic mechanisms. The secular
evolution of all these determine collectively whether and when
conditions for the astrobiological phase transition of the Galaxy
will be satisfied. (Of course, if GRBs are the most important
physical mechanism of extinction, as Annis suggested, than their
distribution function will dominate the global risk function and
force the phase transition.) GRB regulation has an important
correlation property: the rhythm of biological extinctions should
be synchronized (up to the timescales of transport times $\sim 10^4$ yrs
for $\gamma$-rays and high-energy cosmic rays) in at least part of the
histories of all potentially habitable planets. In fact, a bold
hypothesis has been put forward recently by Melott et al. (2004)
that a known terrestrial mass extinction episode, one of the "Big
Five" (the late-Ordovician extinction, cca. 440 Myr before
present), corresponds to a Galactic GRB event.

It is intuitively clear that such correlated behavior undermines
Carter's argument. With a set of modest additional assumptions it
is possible to show it quantitatively. For instance, in Figures 1-4
we show results of a simple numerical experiments performed in
order to see how timescale forcing arises in simplified evolving
systems.  This presents a simple realization of the
astrobiological regulation model of Annis (1999). GRBs are taken
to be random events occuring with exponentially decreasing
frequency
\begin{equation}
\label{jedan} \nu (t) = \nu_0 \exp \left( - \frac{t}{t_\gamma}
\right) ,
\end{equation}
with the fixed characteristic timescale  Gyr in accordance with
the cosmological observations (e.g., Bromm and Loeb 2002), and
biological timescales for noogenesis are random sample from a
uniform distribution between $10^8$ (minimum suggested by McKay
1996) and $10^{16}$ yrs  (the total lifetime of the Galaxy as a well
defined entity; Adams \& Laughlin 1997). It has
been assumed
that the ages of planets are distributed according to the Lineweaver (2001)
age-distribution for terrestrial planets and that the GRBs occur  along the whole timespan considered in Lineweaver (2001). It is taken that the chain
of events leading to life and intelligence can be cut by a
sufficiently strong environmental perturbation at any planet in our toy-model Galaxy with
probability Q and its astrobiological clock reset; this includes cases in
which the destruction of local biospheres is not complete, but the outcome is
sufficiently deflected from the pathways leading to noogenesis that the "new"
timescale shoots out of our temporal window. If the "new" timescale falls within
the temporal window (corresponding to more than a single noogenesis per planet)
it is counted. Thus, the toy model
counts only planets achieving noogenesis (emergence of intelligent observers)
at least once and it
does not take into account any subsequent destructive processes,
either natural or intelligence-caused (like nuclear or biotech
self-destruction). Probability $Q$ can be regarded as
both (1) a geometrical probability of an average habitable
planet being in the "lethal zone" of a GRB, and (2) probability describing
more complex effects dealing with the physics and ecology of the extinction
mechanism. It is important to keep in mind that both these effects can be subsumed
into a single quantity in simple models, but more sophisticated future work
will include two probability parameters.

\section{3. CARTER'S ARGUMENT}

The well-known argument against the existence of extraterrestrial
intelligence (henceforth ETI) due to the astrophysicist Brandon
Carter (1983), and developed by various authors (e.g., Barrow and
Tipler 1986), goes as follows. If astrophysical ($t_*$) and
biological ($t_b$) timescales are truly uncorrelated, life in
general and intelligent life in particular forms at random epoch
with respect to the characteristic timescale of its astrophysical
environment (notably, the Main-Sequence lifetime of the
considered star). In the Solar system, $t_* \simeq t_b$, within the factor
of two. However, in general, it should be either $t_b >> t_*$ or $t_b \simeq
t_*$ or $t_* >> t_b$. The second case is much less probable a priori
in light of independent nature of these quantities. Carter
dismisses the third option either, since in that case it is
difficult to understand why the very first inhabited planetary
system (that is, the Solar System) exhibits $t_* \simeq t_b$ behaviour. On
the contrary, we would then expect that life (and intelligence)
arose on Earth, and probably at other places in the Solar System,
much earlier than they in fact did. This gives us probabilistic
reason to believe that $t_b >> t_*$ (in which case the observation
selection effect explains very well why we do perceive the $t_*
\simeq t_b$ case in the Solar System). Thus, the extraterrestrial life and
intelligence have to be very rare, which is the reason why we
have not observed them so far, in spite of the conjecture that
favorable conditions for it exist at many places throughout the
Galaxy.

}
\end{multicols}

\centerline{\includegraphics[height=8cm]{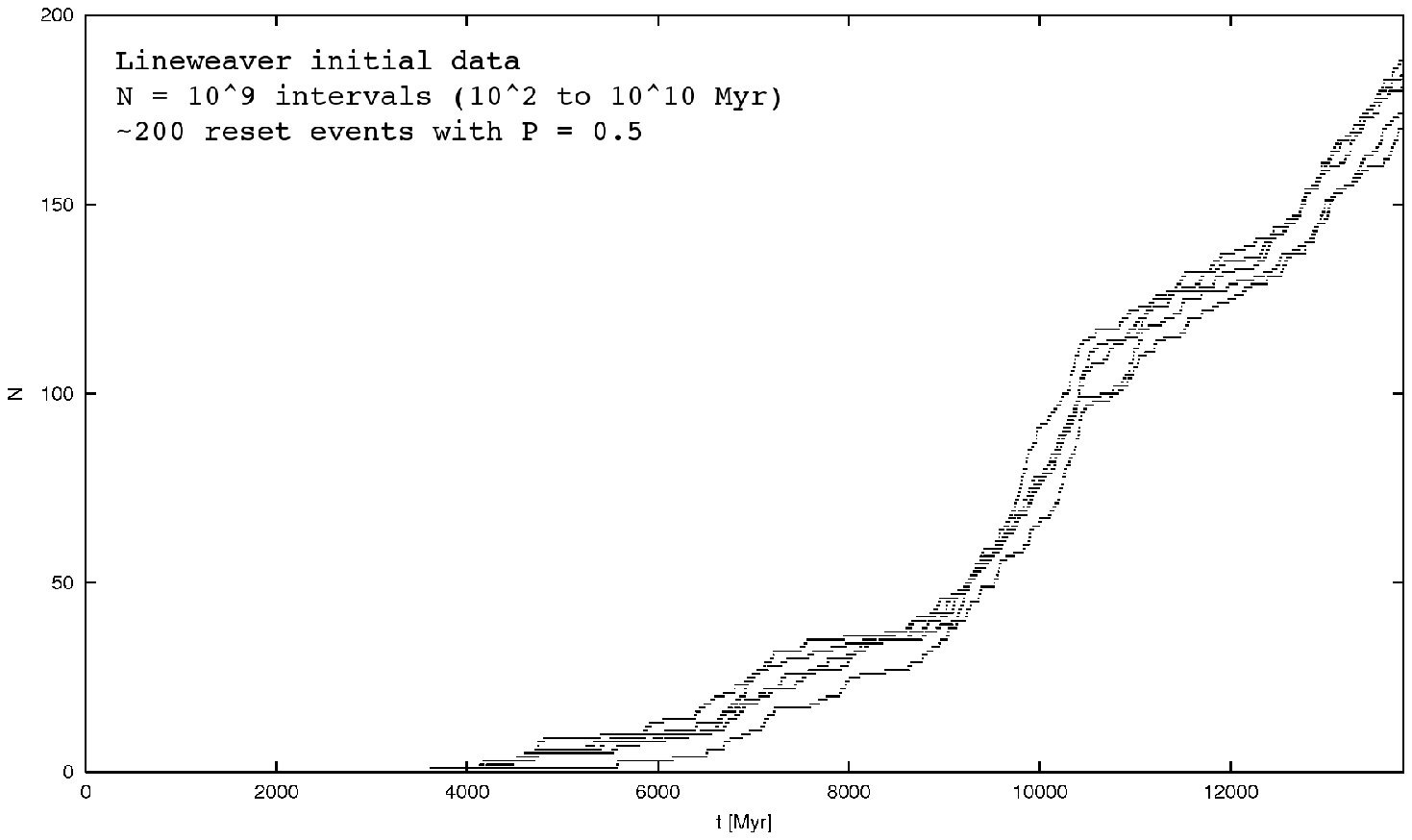}}
\figurecaption{1.}{Five runs of the simplest model with exponentially distributed GRBs, Lineweaver age distribution, and probability of sufficiently strong perturbation equals to $Q=0.5$.}

\centerline{\includegraphics[height=8cm]{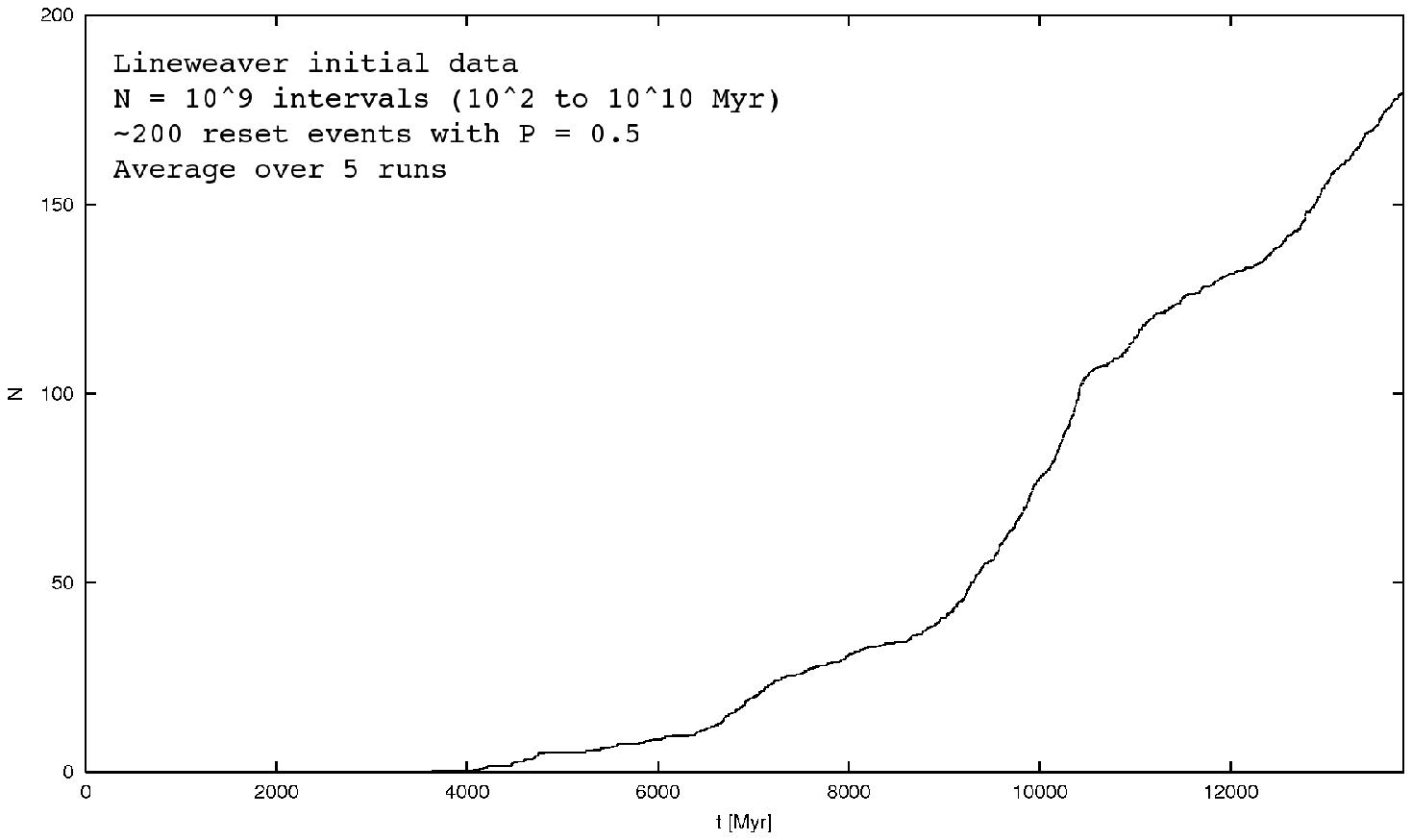}}\figurecaption{2.}{The mean value of runs shown in the previous figure for $Q=0.5$.}

 \centerline{\includegraphics[height=8cm]{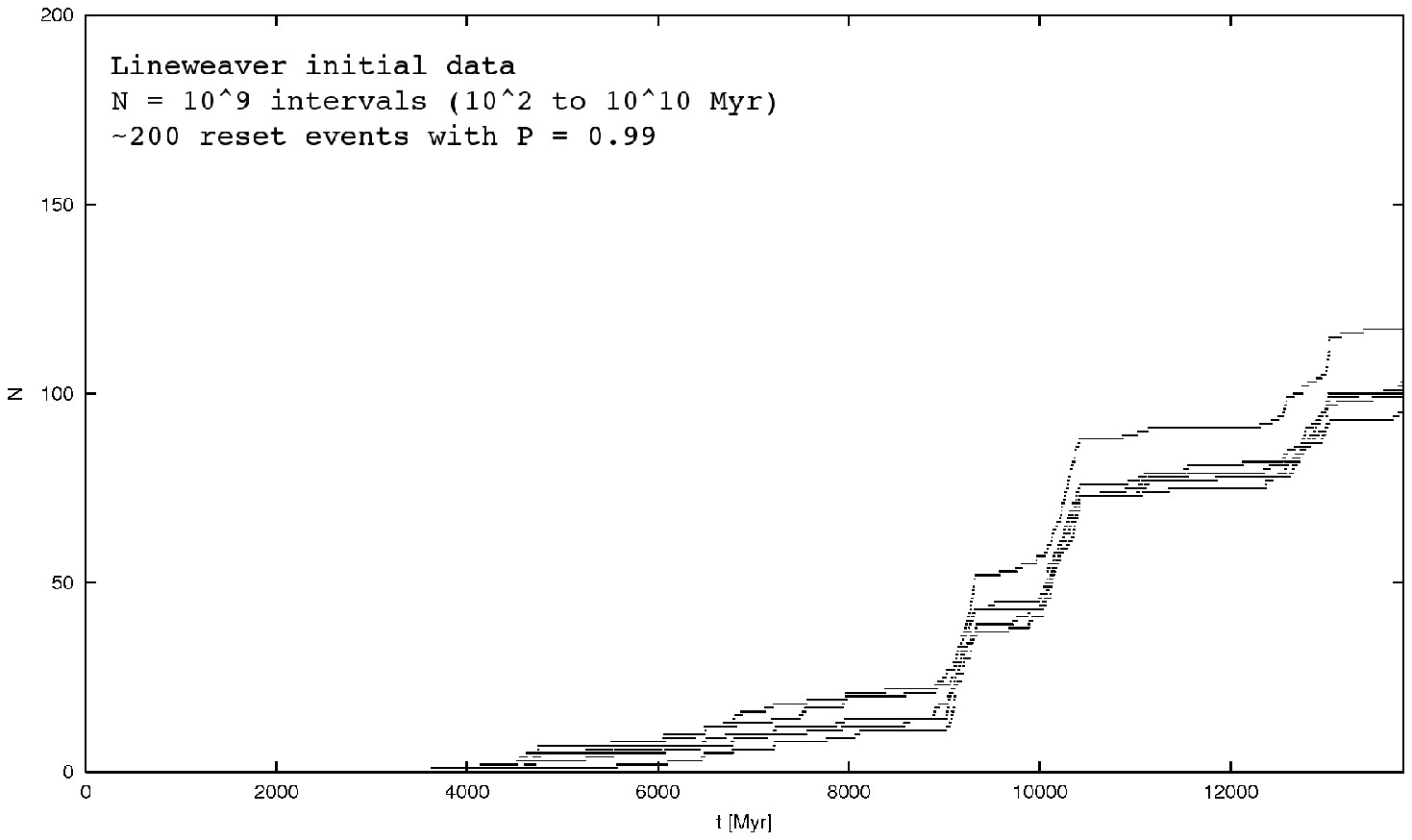}}\figurecaption{3.}{Five runs of the simplest model
with exponentially distributed GRBs, Lineweaver age distribution, and
probability of sufficiently strong
perturbation equals to $Q=0.99$.}

\centerline{\includegraphics[height=8cm]{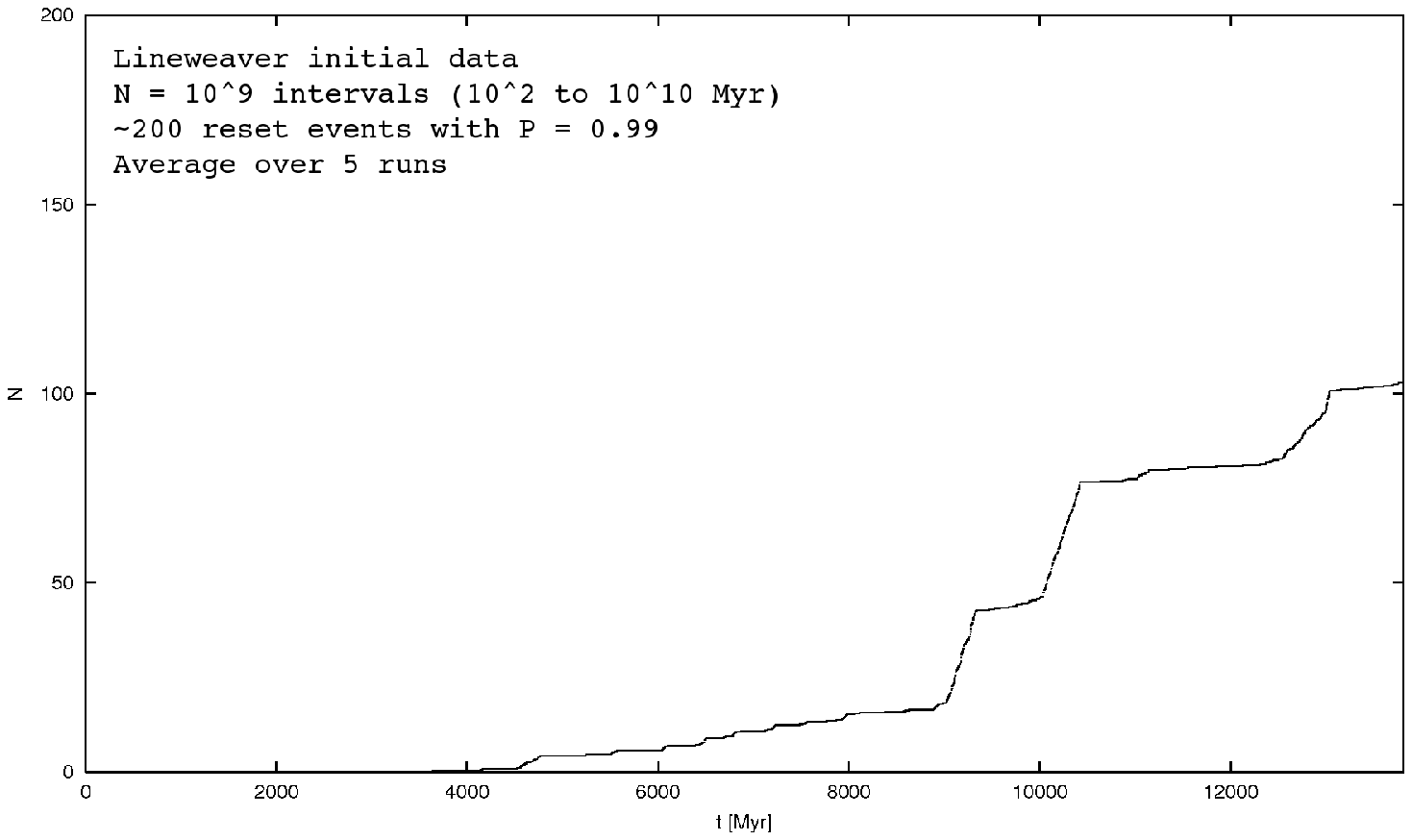}} \figurecaption{4.}{The mean value of runs shown in
the previous figure for $Q=0.99$.}

\begin{multicols}{2}
{

It is clear that the conclusion of Carter's argument depends on the
validity of the independence assumption. In the first place, it is the independence
of biological and astrophysical processes, but the two are linked through
further assumption of the independence of individual sites of biogenesis and
noogenesis (namely, individual planetary systems containing Earth-like planets).
We can clearly undermine this assumption by showing that the catastrophic events,
like GRBs, which influence a large part or all of GHZ {\em induce temporal
correlations\/} between astrobiological histories of these sites.

The conclusions one can draw from our simple models is that, for sufficiently
destructive regulation events in general and GRBs in particular, the timescale
forcing occurs in the system, and the assumption of independence fails. This effect
is particularly visible in the Figure 4, where long plateaus (incidentally, one
including the measured age of the Solar System!) are clearly visible, i.e. a
significant number of habitable planets have their timescales correlated in this manner.
This significantly reduces the rationale behind Carter's argument.

\section{4. DISCUSSION AND FUTURE PLANS}

In the work of Livio (Livio 1999, and references therein) the
author implies that the independence assumption can be undermined
by noticing that circumstellar habitable zones exists only around
stars in the spectral range of about F5 to mid-K and that the
buildup of oxygen in planetary atmospheres and the ozone layer
formation provide a mechanism for linking astrophysical properties
with the timescales for biological evolution. Livio's model has
the lower limit for $t_*$ of approximately 3 Gyr (the main
sequence lifetime of F5 stars). However, it has significant
limitations; notably, it takes into account only biospheres very
similar to Earth's and neglects, for instance, possibility of
habitable moons around Jovian planets (e.g.\ Williams, Kasting \&
Wade 1997). In general, though, our simple model is in rough
agreement with Livios findings. According to our results it is to
be expected that in the "near" future even larger values of $t_b$
fall through the temporal window.

We conclude that even the simplest preliminary models show it is too
early to draw sceptical conclusions
about the abundance of extraterrestrial life and intelligence
from our single data point via the "anthropic" argument of Carter
(1983). In addition to other deficiencies of the argument pointed
out in the literature, we emphasize that a picture in which
regulation mechanism(s) reset local astrobiological clocks (which,
consequently, tick rather unevenly) offers a way to reconcile
both our astrophysical knowledge and the idea about multiple
habitats of life and intelligence in the Galaxy. However, they are very
unevenly distributed in the course of GHZ history. In other words,
Earth may be rare in time, not in space! Quite contrary to the
conventional wisdom, we should not be surprised if we encounter
many "Earths" throughout the Galaxy {\em at this particular epoch},
at stages of evolution of their biospheres similar to the
one reached at Earth, or at least differing for a time factor much
less than the Gyr-scale one obtained by straightforward projecting
Lineweaver's distribution. Unsupported assumption of gradualism is
identified as the main source of confusion and unwarranted SETI
skepticism (for a related discussion in the context of
the Drake equation see Cirkovic 2004). In particular, we expect that
future more detailed models will be able to refine these results, to show
relative importance of various local and global effects in the resulting
noogenesis timescales, and to point a way toward better understanding
the observed "Great Silence" and the likely SETI targets.

\vspace{1cm}

\noindent {\bf Acknowledgements.} This project has been supported
by the Ministry of Science of the Republic of Serbia through the
project No. 146012, "Gaseous and stellar components of galaxies:
interaction and evolution".

\vspace{0.4cm}

\references

Abbas, S. and Abbas, A.: 1998, \journal{Astropart. Phys.}, \vol{8}, 317-320.

Adams, F. C. and Laughlin, G.: 1997, \journal{Rev.\ Mod.\ Phys.},
\vol{69}, 337.

Annis, J.: 1999, \journal{J. Brit. Interplan. Soc.}, \vol{52}, 19-22. (preprint
astro-ph/9901322).

Asher, D. J., Clube, S. V. M., Napier, W. M. and Steel, D. I.:
1994, \journal{Vistas in Astronomy},  \vol{38}, 1-27.

Barrow, J. D. and Tipler, F. J.: 1986, Cosmological Principle (New York: Oxford University Press).

Brin, G. D.: 1983, \journal{Royal Astron. Soc. Quart. Jrn.}, \vol{24}, 283-309.

Bromm, V. and Loeb, A.: 2002, \journal{Astrophys. J.}, \vol{575}, 111-116.

Carter, B.: 1983, \journal{Philos. Trans. R. Soc.
London A}, \vol{310}, 347-363.

\'Cirkovi\'c, M. M.: 2004, \journal{Astrobiology}, \vol{4}, 225-231.

Collar, J. I.: 1996, \journal{Phys. Rev. Lett.}, \vol{76},   999-1002.

Dar, A.: 1997, "Life Extinctions by Neutron Star Mergers,"
in Very High Energy Phenomena in the Universe, Morion Workshop,
ed. by Y. Giraud-Heraud and J. Tran Thanh Van (Editions
Frontieres, Paris), 379-386.

Duric, N. and Field, L.: 2003, \journal{Serb. Astron. J.}, \vol{167},
1-11.

Leitch, E. M. and Vasisht, G.: 1998, \journal{New Astronomy}, \vol{3}, 51-56.

Lineweaver, C. H.: 2001, \journal{Icarus}, \vol{151}, 307-313.

Lineweaver, C. H., Fenner, Y. and Gibson, B. K.: 2004, \journal{Science}, \vol{303}, 59-62.

Livio, M.: 1999, \journal{Astrophys. J.}, \vol{511},
429-431.

McKay, C. P.: 1996, Time for intelligence on other planets, in:
Circumstellar Habitable Zones, Proceedings of The First
International Conference, ed. L. R. Doyle, Travis House
Publications, Menlo Park, pp. 405-419.

Melott, A. L. et al.: 2004, \journal{Int. J. Astrobiol.}, \vol{3}, 55-61.

Rampino, M. R.: 1997,  \journal{Cel. Mech.   and Dyn. Astron.}, \vol{69}, 49-58.

Scalo, J. and Wheeler, J. C.: 2002, \journal{Astrophys. J.}, \vol{566}, 723-737.

Shaviv, N. J.: 2002, \journal{New Astronomy}, \vol{8},
39-77.

Thomas, B. C., Jackman, C. H., Melott, A. L., Laird, C. M.,
Stolarski, R. S., Gehrels, N., Cannizzo, J. K. and Hogan, D. P.:
2005, \journal{Astrophys. J.}, \vol{622}, L153-L156.

Thorsett, S. E.: 1995, \journal{Astrophys. J.}, \vol{444}, L53-L55.

Williams, D. M., Kasting, J. F. and Wade, R. A.: 1997,
\journal{Nature}, \vol{385}, 234-236.

\endreferences

}
\end{multicols}
\vfill\eject

{\ }

% Serbian abstract

% Title

\naslov{O PRINUDNIM KORELACIJAMA VREMENSKE SKALE U ASTROBIOLOGIJI}

% Authors

\authors{B. Vukoti{\' c}\footnote[1]{Prepiska: {\tt
bvukotic@aob.bg.ac.yu}.} and M. M. {\' C}irkovi{\' c}}
\vskip.5mm

% Address
\address{Astronomical Observatory Belgrade,\break Volgina 7, 11160 Belgrade-74, Serbia}

\vskip.7cm

% UDC

\centerline{UDK \udc}

% Papertype

\centerline{\rit Preliminarni izve{\ss}taj}

\vskip.7cm

\begin{multicols}{2}
{

% Abstract

\rrm Istra\zz ujemo efekte korelisanih globalnih mehanizama regulacije, posebno Galakti\ch kih gama b\lj eskova (GB), na vremensku raspodelu hipoteti\ch kih nase\lj enih planeta, upotreb\lj avaju\cc i jednostavne Monte Karlo numeri\ch ke eksperimente. Polaze\cc i od nedavno dobijenih modela starosti planeta u Galakti\ch koj Nasta\nj ivoj Zoni (GNZ), dobijamo da su vremena neophodna za biolo\ss ku evoluciju na nasta\nj ivim planetama Mle\ch nog Puta visoko korelisana. Ovi rezultati idu suprotno poznatom anti-SETI antropi\ch kom argumentu Kartera, i daju nerazra\dj enu podr\ss ku za trenutne i budu\cc e projekte SETI posmatra\nj a.
}
\end{multicols}

\end{document}